\documentclass{elsart}
\usepackage{graphicx}

\begin{document}
\begin{frontmatter}

\title{Dominant $(g_{9/2})^2$ neutron configuration in the $4^+_1$ state of  
$^{68}$Zn  based on new $g$ factor measurements}

\author[bonn]{J.~Leske},
\author[bonn]{K.-H.~Speidel},
\author[bonn]{S. Schielke},
\author[strasbourg]{J.~Gerber},
\author[munich]{P.~Maier-Komor},
\author[oslo]{T.~Engeland},
\author[oslo]{M.~Hjorth-Jensen}

\address[bonn]{Helmholtz-Institut f\"ur Strahlen und Kernphysik, 
Universit\"at Bonn, Nussallee 14-16, D-53115 Bonn, Germany}

\address[strasbourg]{Institut de Recherches Subatomiques, F-67037 Strasbourg, France}

\address[munich]{Physik-Department, Technische Universit\"at M\"unchen, James-Franck-Str., 
D-85748 Garching, Germany}

\address[oslo]{Department of Physics and Center of Mathematics for Applications, 
University of Oslo, N-0316 Oslo, Norway}


\date{\today}

\begin{abstract}\\
The $g$ factor of the $4_1^+$ state in $^{68}$Zn has been remeasured with improved 
energy resolution of the detectors used. The value obtained is consistent 
with the previous result of a negative $g$ factor thus confirming the 
dominant $0g_{9/2}$  neutron nature of the $4_1^+$ state. In addition, the accuracy 
of the $g$ factors of the $2_1^+$, $2_2^+$ and $3_1^-$  states has been improved and 
their lifetimes were well reproduced. New large-scale shell model calculations 
based on a $^{56}$Ni core and an $0f_{5/2}1pg_{9/2}$ 
model space yield a theoretical value, $g(4_1^+) = +0.008$. Although the calculated value is small, it 
cannot fully explain the experimental value, 
$g(4_1^+) = -0.37(17)$. The magnitude of the deduced B(E2) of the $4_1^+$ and
 $2_1^+$ 
transition is, however, rather well described. 
These results demonstrate 
again the importance of $g$ factor measurements for nuclear structure determinations 
due to their specific sensitivity to detailed proton and neutron components 
in the nuclear wave functions.
\end{abstract}


\end{frontmatter}

\section{Introduction}
Measurements of nuclear magnetic dipole and electric quadrupole moments  
provide valuable insights into the nuclear structure based on the intriguing 
interplay of single particle and collective degrees of freedom. In particular, 
nuclei in the $sd$- and $fp$-shell model space have recently received considerable 
attention through many new experiments inspired by the unique possibility 
to compare extensive data with results from large-scale shell model calculations. 
This progress is also related with the observation that general features of the nuclear 
medium can be studied in lighter nuclei as well. For instance, superdeformation 
which had been exclusively investigated for a long time on nuclei in the 
rare earth mass region and beyond \cite{jk1999} has recently been 
observed in $^{36}$Ar \cite{so2004}, $^{40}$Ca \cite{id2001}, $^{60}$Zn \cite{sv1999}  
and $^{62}$Zn \cite{sv1997}. The obvious advantage 
in all these cases is the possibility to directly compare data with 
structure calculations based on highly developed nuclear shell model 
codes thus providing new features of this phenomenon and collectivity in 
general on a microscopic non-phenomenological level.\\
\begin{figure}
\begin{center}
\resizebox{5,7cm}{7cm}
      {\includegraphics{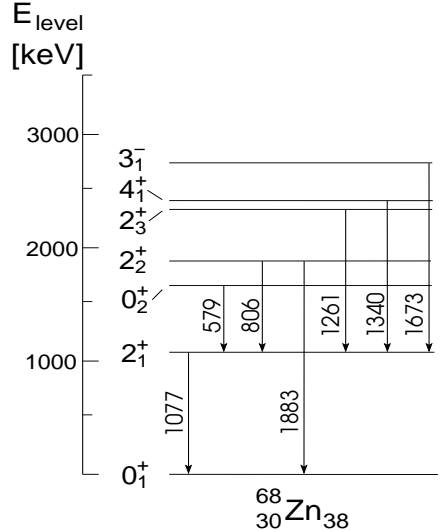}}
\caption{Level scheme of $^{68}Zn$ with $\gamma$ transitions relevant to the present work.}
\label{gr:level}
\end{center}
\end{figure}
In the same context measurements of $g$ factors and B(E2) values have 
been carried out for even-A Zn isotopes with  $A=62-70$ \cite{kenn2002}. Two sets of 
large-scale shell model calculations were applied, based either on a $^{40}$Ca core 
and an $fp$-shell model space or a $^{56}$Ni core and an $fpg$-shell configuration 
space with the inclusion of the $0g_{9/2}$ orbital; the latter was considered to 
be particularly important for the heavier Zn isotopes. In these 
measurements Coulomb excitation of 160 MeV Zn projectiles was achieved 
in collisions with a carbon target employing inverse kinematics. At this beam 
energy essentially the first $2^+$ states were strongly excited whereas 
higher-lying states were only weakly populated.\\
\indent This deficiency was overcome in a succeeding experiment aiming at 
higher excitation energies of both $^{64}$Zn and $^{68}$Zn by increasing the projectile 
energies to 180 MeV close to the Coulomb barrier \cite{leske2005}. In these new measurements 
the $g$ factors and the B(E2) values of the now accessible $4_1^+$ states were 
determined for the first time. Whereas the g($4_1^+$) value of $^{64}$Zn was found to 
be consistent with predictions of both the collective and the spherical shell model, 
the value for $^{68}$Zn of g($4_1^+$) = $-0.4(2)$ turned out to be in severe conflict with 
both model predictions. The negative sign of the $g$ factor was a clear indication 
for a dominant  $0g_{9/2}$ neutron component in the wave function, whereby the 
Schmidt value is $g_{\mathrm{Schmidt}} = -0.467$ which, however, was not verified by the 
calculations.  On the other hand,  the newly determined B(E2) values were very 
well explained by these shell model calculations based on a $^{56}$Ni core 
plus $0g_{9/2}$ neutrons. Evidently, in this case the explicit inclusion of the 
$0g_{9/2}$ orbital seems to be crucial for $^{68}$Zn, as the alternative calculations, 
assuming an inert $^{40}$Ca core but excluding the $0g_{9/2}$ orbital, underestimate the experimental E2 strength.	
	
\begin{figure} 
\begin{center}
\resizebox{\columnwidth}{!}
      {\includegraphics{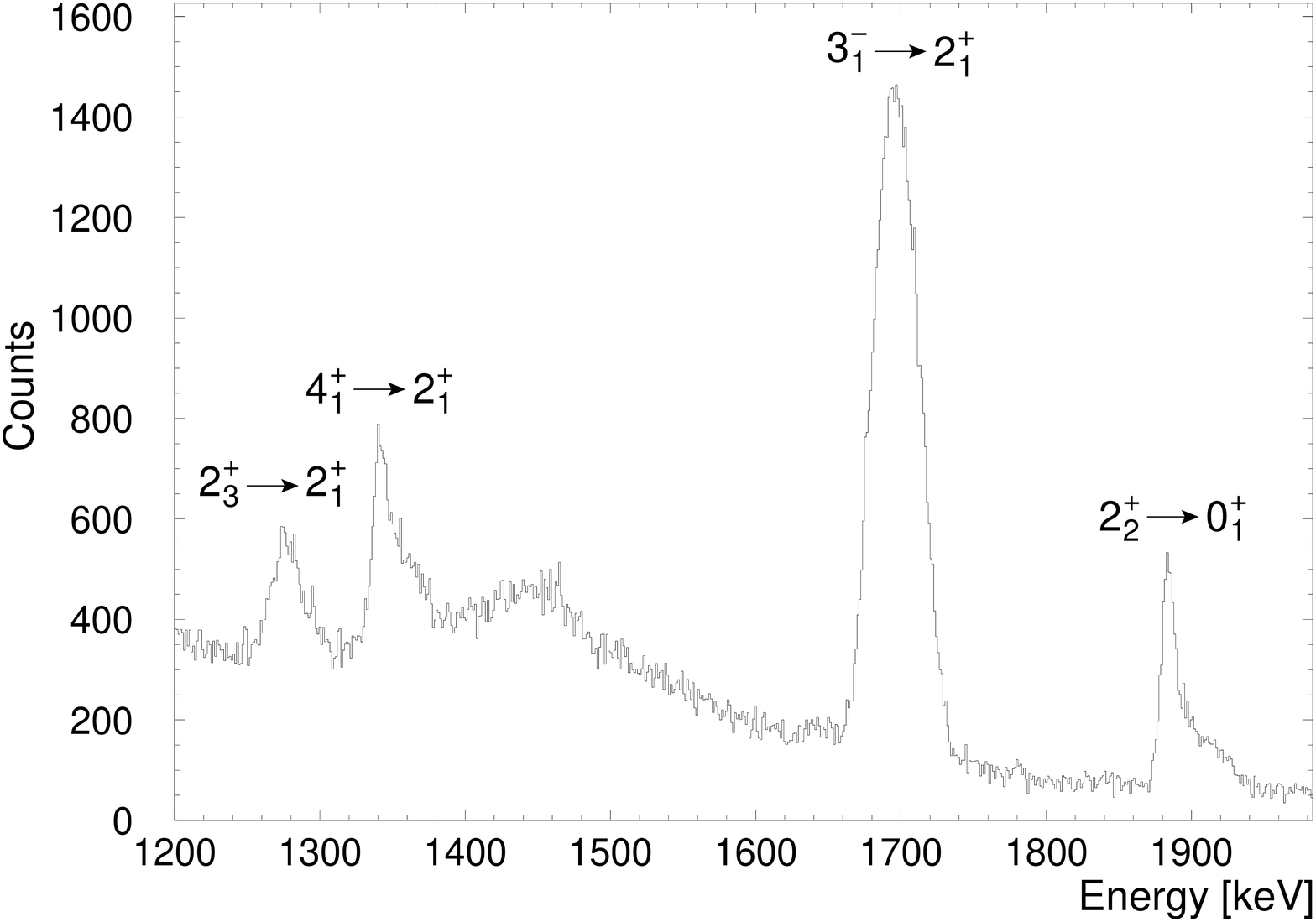}}
\caption{Relevant $\gamma$-coincidence spectrum of a Ge detector placed at $\Theta_{\gamma}$=$65^{\circ}$ relative to the beam axis. The Doppler-broadened lineshapes reflect the nuclear lifetimes.}
\label{gr:ge_spec}
\end{center}
\end{figure}
In order to set the $g$ factor result of the $^{68}$Zn($4_1^+$) state on more solid grounds 
new measurements have been performed. As emphasized in~\cite{leske2005} such an 
effort is required as in the previous measurements the energy resolution 
of the NaI(Tl) scintillators used for $\gamma$ detection did not allow to 
fully separate the relevant ($4_1^+ \rightarrow 2_1^+$)  
$\gamma$ line from a neighbouring ($2_3^+ \rightarrow 2_1^+$) 
1261 keV line. As the latter was also strongly Doppler-shifted 
due to the short nuclear lifetime, the separation of the two $\gamma$ lines was 
particularly crucial for the detector pair placed in the forward hemisphere. 
On the other hand, estimates of an eventual admixture of the $2_3^+$ state 
to the measured precession of the $4_1^+$ state, based on the observed 
line intensities and the angular correlations of the corresponding 
$\gamma$ transitions, could not explain the deduced  $g$ factor, even under the 
assumption of a negative $g$ value for the $2_3^+$  state. 

\section{Experimental details}

In the present experiment a beam of isotopically pure $^{68}$Zn ions was 
accelerated to an energy of 180 MeV at the Munich tandem accelerator 
providing intensities of  ~20 $e$nA on a multilayered target. The 
latter consisted of 0.44 mg/cm$^2$ natural carbon deposited on a 3.34 mg/cm$^2$ Gd 
layer, 
which was evaporated on a 1.4 mg/cm$^2$ Ta foil, 
backed by a 4.49 mg/cm$^2$ Cu layer. Between C and Gd as well as between Ta and Cu 
thin layers of natural titanium ($\sim 0.005$ mg/cm$^2$) provided good adherence 
being very crucial for the precession experiments. The same target had been 
used in former measurements under almost identical conditions \cite{leske2005}. The target 
was cooled to liquid nitrogen temperature and magnetized to saturation 
by an external field of 0.06 T. The relevant level scheme of $^{68}$Zn for 
Coulomb excitation is shown in Fig. 1 \cite{js2000}. The excited Zn nuclei move 
through the Gd layer at mean velocities of $\sim 5.9 v_0$ ($v_0 = e^2/\hbar$) experiencing 
spin precessions in the transient field and are ultimately stopped 
in the hyperfine-interaction-free environment of the Cu backing.
	
The de-excitation $\gamma$ rays were measured in coincidence with the 
forward scattered carbon ions detected in a 100 $\mu$m Si detector at $0^{\circ}$. A 5 $\mu$m 
thick Ta foil between target and particle detector served as a beam stopper 
which, however, was transparent to the carbon recoils. As in the previous 
experiments the detector was operated at a very low bias 
of  $\sim 5$ V to establish a thin depletion layer for separating 
the energies of carbon ions from those of light particles 
as protons and $\alpha$ particles resulting from sub-Coulomb fusion 
and transfer reactions. Under these conditions very clean 
$\gamma$-coincidence spectra were obtained. Intrinsic Ge detectors of  $\sim 40 \%$ relative 
efficiency were used for $\gamma$ detection. Fig. 2 shows a typical coincidence 
spectrum with emphasis on the ($4_1^+ \rightarrow 2_1^+$) $\gamma$ line. 
Evidently, all relevant $\gamma$ lines were well resolved allowing a rigorous 
determination of their intensities required for the angular correlations 
as well as for the precessions of the nuclear states in question.

	Particle-$\gamma$ angular correlations $W(\Theta_{\gamma})$ have been measured for 
determining the slope $\mid S \mid$ = [1/W($\Theta_{\gamma}$)]$\cdot$[dW($\Theta_{\gamma}$)/d$\Theta_{\gamma}$] in the rest frame 
of the $\gamma$ emitting nuclei at $\Theta_{\gamma}^{lab}$ = $\pm 65^{\circ}$ and $\pm 115^{\circ}$, where the sensitivity 
to the precessions was optimal for all transitions of interest. 
Precession angles, $\Phi^{exp}$, were derived from counting-rate ratios 
'R' for 'up' and 'down' directions of the external magnetizing field 
which can be expressed as \cite{leske2005} ,
\begin{eqnarray}
\Phi^{exp}=\frac{1}{S} \cdot \frac{\sqrt{R}-1}{\sqrt{R}+1}=g
\frac{\mu_N}{\hbar}
\int^{t_{out}}_{t_{in}}{B_{TF}(v_{ion}(t))e^{-\frac{t}{\tau}}dt}
\label{eq:phi}
\end{eqnarray}
where $g$ is the $g$ factor of the nuclear state and $B_{TF}$ the transient 
field acting on the nucleus during the time interval 
$(t_{\mathrm{out}} - t_{\mathrm{in}}$) which 
the ions spend in the gadolinium layer of the target; 
the exponential accounts for nuclear decay with lifetime $\tau$ in the Gd layer. 
	
Simultaneously with the precessions the lifetimes of 
several excited states have been redetermined using the 
Doppler-Shift-Attenuation-Method (DSAM). For the analysis of the 
Doppler-broadened lineshapes, which were observed with a Ge detector 
placed at $0^{\circ}$ to the beam direction, the computer code LINESHAPE \cite{wj1994} 
has been used. 
Specific details of the analysis procedure are given in \cite{kenn2002,leske2005}.

\section{Results and discussion}

The $g$ factors have been determined from the measured precession angles 
by calculating the effective transient field $B_{TF}$ on the basis of 
the empirical linear parametrization (see \cite{kenn2002,leske2005}):
\begin{equation}
B_{\mathrm{TF}}(v_{\mathrm{ion}}) = G_{\mathrm{beam}} \cdot B_{\mathrm{lin}},
\end{equation}
with
\begin{equation}
B_{\mathrm{lin}} = a(Gd) \cdot Z_{\mathrm{ion}} \cdot v_{\mathrm{ion}}/v_0,
\end{equation}
where the strength parameter $a(Gd) = 17(1) T$ \cite{leske2005}, and 
$G_{\mathrm{beam}} = 0.61(6)$ 
is the attenuation factor accounting for the demagnetization 
of the Gd layer induced by the Zn beam (see \cite{leske2005}). The same scheme 
has been successfully applied in many former measurements. 
\begin{table*}[t]
\begin{center}
\caption{Summary of the measured slopes of the angular correlations at $\Theta_{\gamma}^{Lab}$=$\pm65^{\circ}$, precession angles and lifetimes of excited states in $^{68}Zn$. The $\Phi_{lin}/g$ values were calculated using Eqs.(1)-(3). The $g$ factors deduced and the newly determined lifetimes are compared with previous results~\cite{leske2005}.}
\begin{tabular}{c|c|cc|c|c|c|cc}
 ~~$E_{x}$[MeV]~~ & ~~$ I^{\Pi}$~~ & \multicolumn{2}{c|}{$\tau$[ps]} & ~~$\vert S(65^{\circ}) \vert$ ~~&~~ $\Phi^{exp}$~~&~~ $\Phi^{lin}/g$ ~~&\multicolumn{2}{c}{$g(I)$}\\
  & & ~~~present~~~ &~~~~\cite{leske2005}~~~~&~~~$[mrad]^{-1}$~~~&~~~[mrad]~~~&~~~[mrad]~~~&~~~present~~~&~~~\cite{leske2005}~~~\\
[2mm]\hline
  1.077 & $2_1^+$ & 2.34(4) & 2.32(5) & 2.133(15) & 15.2(2) & 26.4(26) & +0.58(6) & +0.505(38)\\[2mm]
  1.883 & $2_2^+$ & 1.5(1) & 1.4(1) & 1.24(16) & 12(4) & 24.9(25) & +0.48(17) & +0.53(15) \\[2mm]
  2.338 & $2_3^+$ & 0.47(6) & 0.45(4) & -- & -- & -- & -- & -- \\ [2mm]
  2.417 & $4_1^+$ & 1.18(8) & 1.10(8) & 0.85(18) & $-6(8)$ & 23.8(24) & $-0.3(3)$ & $-0.4(2)$ \\
  2.751 & $3_1^-$ & 0.37(1) & 0.38(2) & 0.310(52) & 5(6) & 16.7(17) & +0.3(4) & +0.4(3) \\ [1mm]
  
\end{tabular}
\label{tb:table1}
\end{center}
\end{table*}
	Precession and lifetime data from a single run are summarized 
in Table~\ref{tb:table1} together with the deduced $g$ factors which 
are compared with previous results \cite{leske2005}. Evidently, all newly determined 
lifetimes and $g$ factors are in good agreement with earlier data \cite{leske2005}. 
Furthermore, the $g$ factor of the $4_1^+$ state with its negative sign is confirmed 
whereby the relatively large error in the magnitude is of purely statistical 
origin due to the small excitation cross-section of the nuclear state 
and the low $\gamma$-detection efficiency of the Ge 
detectors. The striking difference in the $g$ factors of the $4_1^+$ states between $^{64}$Zn and $^{68}$Zn 
is shown in Fig. 3.

	In Table~\ref{tb:table2}, the $g$ factors and the B(E2)'s, both improved in 
accuracy by averaging with the data of \cite{kenn2002,leske2005}, are compared 
with results from large-scale shell model calculations.  

\begin{table*}[t]
\begin{center}
\caption{Comparison of experimental excitation energies, average $g$ factors and $B(E2)$ values (see also~\cite{leske2005}) of $^{64,68}$Zn with results from shell-model calculations (see text for further details).}
\begin{tabular}{c|ccc|c|c|ccc|ccc}
Nucleus & \multicolumn{3}{c|}{$E(I_f^{\pi})$[MeV]} & $I_f^{\pi} \rightarrow I_i^{\pi}$ &~~$\tau(I_f^{\pi})$[ps]~~ &\multicolumn{3}{c|}{$g(I_f^{\pi})$}  & \multicolumn{3}{c}{$B(E2)\downarrow[e^2fm^4]$}\\
 & ~~~exp.~~~ &~~~SM-1~~~&~~~SM-2~~~&   &  &~~~~~~exp.~~~~~&~~~SM-1~~~&~~~SM-2~~~&~~~exp.~~~&~~~SM-1~~~&~~~SM-2~~~ \\[1mm]
\hline
$^{64}$Zn & 0.992 & & 1.057 &~~$2_1^+ \rightarrow 0_1^+$~~& 2.77(6) &+0.446(25)& &+0.813 & 307(7) & & 170.4\\ [2mm]
          & 3.207 & & 2.830 & $4_1^+ \rightarrow 2_1^+$ & 1.12(4) & +0.53(16) & &+1.568 & 185(7) & & 196.1\\ [2mm]
          & 1.799 & & 2.467 & $2_2^+ \rightarrow 0_1^+$ & 2.9(3)  & -- & & +0.334 & 3.5(4) & & 1.2\\ [2mm]\hline
$^{68}$Zn & 1.077 & 0.765 & 0.934 & $2_1^+ \rightarrow 0_1^+$ & 2.33(3) & +0.498(26)& +0.027 & +0.223 & 242(3) & 151.6 & 190.9\\ [2mm]
          & 2.417 & 1.936 & 2.097 & $4_1^+ \rightarrow 2_1^+$ & 1.14(6) & $-0.37(17)$ & +0.066 & +0.008 & 166(9)& 209.2 & 224.6\\ [2mm]
          & 1.883 & 1.318 & 1.629 & $2_2^+ \rightarrow 0_1^+$ & 1.45(7) & +0.51(11) & +0.672 & +0.242 & 15(1) & 26.0 & 3.4\\ [2mm]
\end{tabular}
\label{tb:table2}
\end{center}
\end{table*}

\begin{figure} 
\begin{center}
\resizebox{\columnwidth}{!}
      {\includegraphics{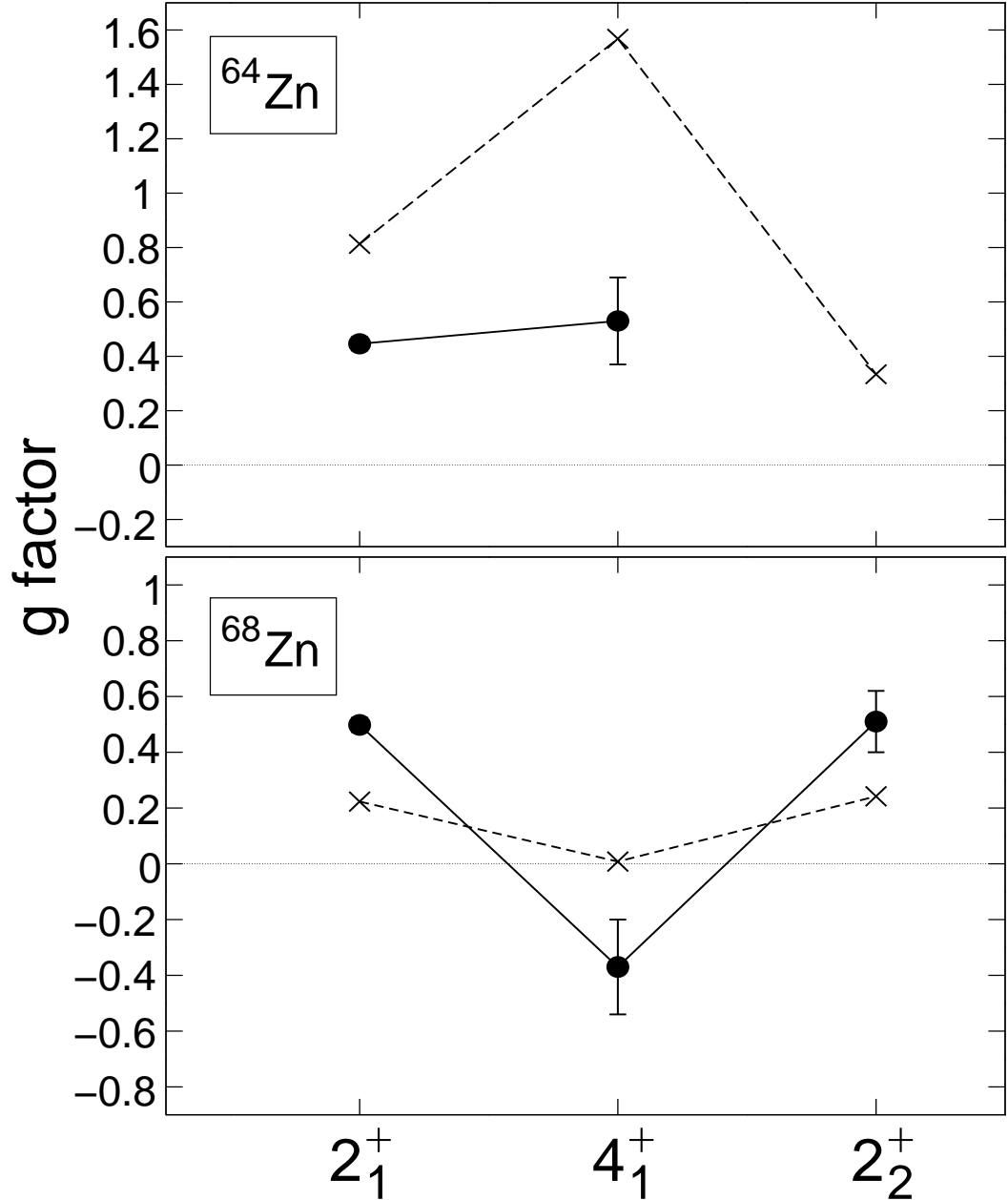}}
\caption{Comparison of experimental $g$ factors associated with the $2_1^+$, $4_1^+$ and $2_2^+$ states in $^{64}Zn$ and $^{68}Zn$ (closed circles) with results from large-scale shell model calculations (crosses, see text). Lines are drawn to guide the eye.}
\label{gr:ge_za}
\end{center}
\end{figure}
In order to study the importance of the $0g_{9/2}$ orbit we have 
performed shell model calculations using $^{56}$Ni as closed shell core, with a model
space defined by protons and neutrons occupying
the single-particle orbitals $0f_{5/2}$, $1p_{3/2}$,
$1p_{1/2}$ and $0g_{9/2}$ ($0f_{5/2}1pg_{9/2}$). To determine the effective interaction we use the recent charge-dependent
potential model of Machleidt, the so-called CD-Bonn interaction \cite{cdbonn}.
The final effective two-body interaction is obtained via many-body perturbation
theory to third order, employing a renormalized nucleon-nucleon interaction 
defined for $^{56}$Ni as closed shell core and 
including folded-diagrams to infinite order. For details, see for example 
Ref.~\cite{hko1995}. A harmonic oscillator basis was used, with an oscillator energy
determined via
$\hbar\Omega = 45A^{-1/3} -25A^{-2/3}$ = 10.1 MeV, A = 56 being
the mass number. For the single-particle energies we employ values
adapted from Grawe in Ref.~\cite{grawe1999}, resulting in 
the energy differences 
$\epsilon_{0g_{9/2}}-\epsilon_{1p_{3/2}}=3.70$ MeV, 
$\epsilon_{0f_{5/2}}-\epsilon_{1p_{3/2}}=0.77$ MeV, 
$\epsilon_{1p_{1/2}}-\epsilon_{1p_{3/2}}=1.11$ MeV for neutrons and
$\epsilon_{0g_{9/2}}-\epsilon_{1p_{3/2}}=3.51$ MeV, 
$\epsilon_{0f_{5/2}}-\epsilon_{1p_{3/2}}=1.03$ MeV, 
$\epsilon_{1p_{1/2}}-\epsilon_{1p_{3/2}}=1.11$ MeV for protons.
The effective interaction has not been corrected for any eventual monopole changes.
This means that the only parameters which enter our calculations are those defining the
nucleon-nucleon interaction fitted to reproduce the scattering data, the experimental
single-particle energies and the oscillator basis. 
Furthermore, for the computation of the B(E2)'s and
$g$ factors, we have used unrenormalized magnetic moments and the canonical
unrenormalized charges for protons and neutrons. The latter entails
charges of $1.5e$ and $0.5e$ for protons and neutrons, respectively. 
The free-nucleon operator for the magnetic moment is defined as
\begin{equation}
\label{eq1f}
{\bf \mu}_{{\rm free}} = g_{l} {\bf l} +
g_{s} {\bf s},
\end{equation}
with $  g_{l}({\rm proton}) = 1.0  $,
$  g_{l}({\rm neutron}) = 0.0  $,
$  g_{s}({\rm proton}) = 5.586  $,
$  g_{s}({\rm neutron}) = -3.826  $.
Note, however, that  the magnetic moment operator in finite nuclei is modified from the
free-nucleon operator
due to core-polarization and meson-exchange
current (MEC) corrections \cite{To87,CT90}.
The effective operator is defined as
\begin{equation}
\label{eq1}
{\bf \mu}_{{\rm eff}} = g_{l,{\rm eff}} {\bf l} +
g_{s,{\rm eff}} {\bf s} +
g_{p,{\rm eff}} [Y_{2} ,{\bf s}] ,
\end{equation}
\noindent where $  g_{x,{\rm eff}} = g_{x} + \delta g_{x}  $, $  x = l
$,
$  s  $ or $  p  $,
with $  g_{x}  $ the free-nucleon, single-particle
$  g  $ factors ($g_p=0$) and $  \delta g_{x}  $ the
calculated correction to it.  Note the presence of a new term
$  [Y_{2} , {\bf s}]  $, absent from the free-nucleon operator, which is
a spherical
harmonic of rank $\lambda'=2$ coupled to a spin operator to form a spherical tensor
of multipolarity $\lambda=1$. In the calculations below we limit ourselves to
a calculation with free operators only.
An obvious modification of the latter is to use renormalized moments, as done
in the recent work of Ref.~\cite{brown2005} for nuclei in the 
$^{132}$Sn mass region.

The results of the calculations for the energies of the
low-lying excited states for both $^{64}$Zn and $^{68}$Zn in Table~\ref{tb:table2}, while 
 $B(E2;2^+_1 \rightarrow 0^+_1$), $B(E2;2^+_2 \rightarrow 0^+_1$) and   
$B(E2;4^+_1 \rightarrow 2^+_1$) and the $g$ factors
$g(2^+_1)$, $g(2_2^+)$  and $g(4^+_1)$ are presented and compared with the available data.  

Two types of shell-model results are discussed and shown in this Table, 
SM-1 and SM-2.
In SM-1 we limit the number of neutrons which can be excited to the $0g_{9/2}$ orbit
to two, whereas SM-2 is the full shell-model calculation. The latter basis is, however,
unnecessarily large since the results are converged with four neutrons at most in
the $0g_{9/2}$ orbit. 
One sees clearly that with at most two neutrons in the $0g_{9/2}$ orbit
the spectrum is rather poorly reproduced.
However, even for the  fully converged calculation 
the first excited  $0^+_2$ state at 1.655 MeV is badly reproduced ($E^{SM-2}$=2.406 MeV), indicating most likely the need
for particle-hole excitations, especially from the $0f_{7/2}$ orbit. 
The typical occupation probabilities in the SM-2 calculation of the  $0g_{9/2}$
neutron single-particle orbit span from 2.2 to 2.4. 
The $1p_{3/2}$
neutron orbit has an average occupancy of 3.5 particles whereas the low-lying 
$1f_{5/2}$ neutron orbit has occupancies around three. 
We note that for the $2^+_1$, $2^+_2$ and the $4^+_1$ states there is a 
satisfactory agreement with the
data. 

For $^{68}$Zn
the B(E2;2$^+_1\rightarrow$0$^+_1$) exhibits a good    
agreement with data using unrenormalized effective charges while the 
B(E2;4$^+_1$ $\rightarrow$2$^+_1$) is overestimated.
This could be ascribed to deficiencies in the two-body Hamiltonian and/or 
omitted degrees of freedom in the model space.
 
For $^{64}$Zn we get 170.4 $e^2$fm$^4$ for the B(E2;2$^+_1\rightarrow$0$^+_1$), to be compared with the experimental value of 307 $e^2$fm$^4$, hinting
at the need of larger effective charges. This means in turn that 
particle-hole excitations involving the $0f_{7/2}$ orbit may be more important for 
$^{64}$Zn than for $^{68}$Zn, in good agreement with previous shell model calculations including
this degree of freedom (see for example the discussions in  
Refs.~\cite{kenn2002,leske2005}).
This is also reflected in the $g$ factors for $^{64}$Zn, 
which tend to be larger than the experimental values.
If we, however, introduce effective magnetic moments by assuming a renormalization factor of $0.7-0.8$ for 
$g_s$ of protons and neutrons, we obtain $g$ factors closer to experiment. 
However, the $0f_{7/2}$ orbit, if coupled
with the $1p_{3/2}$ orbit can yield a negative contribution to the $g$ factors. 
The latter analysis is obviously performed at the level of a simple two-body configuration (see 
for example
discussions in Refs.~\cite{lawson80,arimahorie54,georgiev2001}), however, 
together with the $(0g_{9/2})^2$ configuration these are the only two-body configurations
of interest here which can yield a negative $g$ factor.  
For $^{64}$Zn the occupancy of the 
 $0g_{9/2}$ orbit is less than one, and plays therefore a negligible role in the 
calculation of $g$ factors.
Since excitations from the $^{56}$Ni closed
shell core are not included in the present model space, this reduction cannot be accounted for. 
Thus, the theoretical values
should be larger than the experimental ones.

For $^{68}$Zn and its  $g$ factors we note that for the $2^+$ states there is a fair agreement with
data, confirming the previous shell-model analysis presented in 
Refs.~\cite{kenn2002,leske2005}. For $g(4^+_1)$ we see that there is again a  
change from
the calculation with only two neutrons in the $0g_{9/2}$ orbit to the full calculation.
This displays the role played by the $0g_{9/2}$ orbit, which for the $4^+_1$
has an average occupation probability of 2.4 in the full SM-2 calculation, much larger than we have 
for $^{64}$Zn. 
For SM-1 the corresponding occupation probability is 1.99 and this difference is clearly reflected in the change
of $g(4^+_1)$ from 0.066 to 0.008 demonstrating thereby the dominating role played by the 
$(0g_{9/2})^2$ admixture in the wave function.
However, 
for SM-2 the $g$ factor although very small, is  still positive. 
Introducing effective magnetic moments with a scaling factor of 
$0.7$ reduces the theoretical SM-2 value to $g(4^+_1)=+0.003$. 
We speculate again whether particle-hole excitations 
involving the $0f_{7/2}$ orbit could yield further reductions
and eventually a negative contribution. Furthermore,  
another possibility is to include the effect of meson-exchange currents, as done in
Ref.~\cite{brown2005}. Meson-exchange current corrections arise 
because nucleons in nuclei are
interacting through the exchange of mesons, which can be disturbed
by the electromagnetic field. In terms of effective one-body operators this leads to
a correction term 
$  [Y_{2} , {\bf s}]  $ in Eq.~(\ref{eq1}). In spite of these omitted degrees of freedom,  
we see that our model space displays the 
important role played by the 
$(0g_{9/2})^2$ neutron configuration when going from the SM-1 to the full SM-2 shell-model calculation, 
lending thereby
support to the experimental analysis.  

The role of the neutron $g_{9/2}$ orbit is also seen in the $g$ factors of low-lying $9/2^+$ states in odd $Zn$ isotopes with a large negative value, although smaller in absolute value than the corresponding Schmidt value (see Ref.~\cite{herzog89}).

\section{Summary and conclusions}

We have remeasured 
the $g$ factor of the $4_1^+$ state in $^{68}$Zn, with an improved 
energy resolution of the detectors used. 
The experimental value is  $g(4_1^+) = -0.37(17)$ consistent with our previous measurements.
In addition, the accuracy 
of the $g$ factors of the $2_1^+$, $2_2^+$ and $3_1^-$  states has been improved and 
their lifetimes were well reproduced.
The experimental results for $^{64}$Zn and $^{68}$Zn have been compared with large-scale shell model calculations using
$^{56}$Ni as a closed shell core and a model space consisting of protons and neutrons
occupying the $0f_{5/2}1pg_{9/2}$ single-particle orbits. The agreement between theory and experiment
is rather good and the calculations reproduce well
the experimental trends for the $g$ factors from $^{64}$Zn to $^{68}$Zn, 
although our theoretical approach is not capable of
reproducing the negative sign of the $g(4_1^+)$ value in $^{68}$Zn. This deficiency may be ascribed to particle-hole excitations, with 
the $0f_{7/2}$ orbit playing a major role. 
The effect of meson-exchange currents may also play a role and will be investigated in future works, 
together with the inclusion of the 
$0f_{7/2}$ orbit. In view of the present results similar measurements for the Ge isotones $^{66}$Ge and $^{70}$Ge are highly desirable to search for corresponding effects in the nuclear wave functions.

\begin{ack}
The authors are thankful to the operating staff of the Munich tandem accelerator. Support by the BMBF and the Deutsche Forschungsgemeinschaft is acknowledged.
The work of TE and MHJ has been supported by the Research Council of Norway (Program
for Supercomputing) through a grant of computing time.
Discussions with Alex Brown (MSU) and Georgi Georgiev (CERN) are gratefully acknowledged.
\end{ack}

\end{document}